\documentstyle[twocolumn,prc,aps,epsfig]{revtex}
\input epsf
\newcommand{\be}{\begin{eqnarray}}
\newcommand{\ee}{\end{eqnarray}}
\def\lsim{\mathrel{\rlap{\lower3pt\hbox{\hskip1pt$\sim$}}
     \raise1pt\hbox{$<$}}} 
\def\gsim{\mathrel{\rlap{\lower3pt\hbox{\hskip1pt$\sim$}}
     \raise1pt\hbox{$>$}}} 

\begin{document}

\twocolumn[\hsize\textwidth\columnwidth\hsize\csname
@twocolumnfalse\endcsname

\title{Ampere's Law and Energy Loss in AdS/CFT Duality}

\author{Sang-Jin Sin$^a$ and Ismail Zahed$^b$}
\address {
 $^a$ Department of Physics, Hanyang University, Seoul
133-791, Korea\\
$^b$ Department of Physics and Astronomy\\
State University of New York, Stony Brook, NY 11794-3800}

\date{\today}
\maketitle
\begin{abstract}
We note that the energy loss in ${\cal N}=4$ SYM 
measures directly the spatial string tension 
$\sigma_S=\pi\sqrt{\lambda}T^2/2$ which is at the
origin of the area law for large spatial Wilson
loops. We show that the latter reflects on the
nonperturbative nature of Ampere's law in ${\cal N}=4$
SYM both in vacuum and at finite temperature.

\end{abstract}
\vspace{1.cm}
]
\begin{narrowtext}
\newpage

{\bf 1. Introduction\,\,}
Recent relativistic heavy ion collisions at RHIC
have suggested that the partonic matter is unleashed
in the form of a strongly coupled quark gluon plasma
sQGP which is likely in the liquid phase~\cite{SQGP}.
The key characteristics of this phase reside in its
transport properties and not in its bulk properties.

The transport properties of the sQGP are unknown
from first principles, since most current lattice
formulations are restricted to the Euclidean domain.
Time-like properties of the sQGP are elusive. QCD
just above the critical temperature is in a strongly
coupled Coulomb phase. Recently, it was suggested
\cite{STRONG} that this phase may share similarities
with finite temperature ${\cal N}=4$ SYM in the dual
limit of strong coupling and large $N_c$~\cite{MALDACENA}.

A number of transport properties of the ${\cal N}=4$
SYM theory at finite temperature have been recently
reported~\cite{SON} all consistent with a hydrodynamical
limit in the strongly coupled limit. In~\cite{SIN} we
have emphasized the concept of the energy loss in hot
${\cal N}=4$ SYM theory and its importance for current
RHIC experiments. Indeed, we have shown that any color
triggered wave is absorbed on a distance of $1/\pi T$
irrespective of its frequency/energy content in hot
${\cal N}=4$ SYM. The color opacity length is $1/\pi T$
and short.

Recently, there has been a flurry of activities related to the
issue of absorption or energy loss in hot ${\cal N}=4$ SYM
using heavy probles~\cite{YAFFE,DEREK,GUBSER,HERZOG,CAS,VAZ} and
light dipoles~\cite{URS,URSCRITIC}. One of the central
result is the presence of a dragging force ($\gamma=1/\sqrt{1-v^2}$)

\be
F_x=dE/dx= - \gamma v\, m_T^2,
\ee
proportional to the velocity $v$. Stoke's law hold in hot 
${\cal N}=4$ SYM at strong coupling as first suggested 
in~\cite{GELLMANN} (section Vb).

From the `single string solution' of
\cite{YAFFE,GUBSER} it is clear that the quark-antiquark $Q\overline{Q}$
configuration  with large separation is the mirror reflected
solution of single heavy probe
as shown in Fig.~1. This is in fact supported by the numerical
solution given in~\cite{YAFFE}. In light of this and the well
established fact~\cite{YEE}
that the static color force is screened in hot SYM (see Fig.~1),
it is tempting to conclude that {\it the relevant force in the  medium
is color magnetic rather than color electric.}. Time- and space-like
$Q\overline{Q}$ configurations have been originally analyzed in the
context of scattering at high energy in~\cite{QQ1,QQ2,POLSHINSKI}.

In this short note we argue that this is indeed the case by
discussing ampere's law and the nature of the magnetic force
of hot N=4 SYM both  in the perturbative and AdS/CFT context.
Throughout we will use exchangeably
Euclidean and Minkowski descriptions for the analysis
of the Wilson lines. Which is which will be clear by the
notation and discussion.

\vskip 1.5cm

\begin{figure}[ht]
\begin{center}
\epsfig{figure=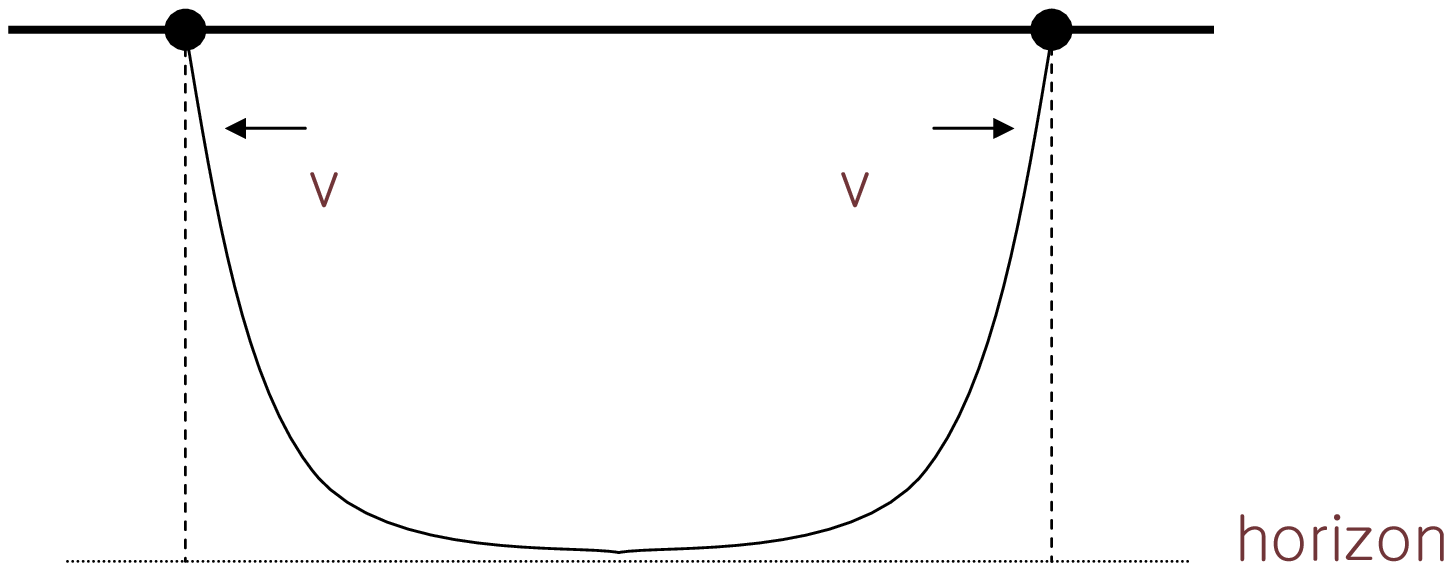,height=3.5cm,width=7cm,angle=0}
\bigskip
\caption{ The string connecting
a receding quark-antiquark with velocity $v$. The dashed lines
are the string configuration for the static case $v=0$.
They show the screened nature of the color
electro-static force.}
\label{wilson}
\end{center}
\end{figure}

\vskip .5cm

{\bf 2. Spatial Wilson Loops\,\,}
At finite temperature {\it parallel} space-like Wilson
loops confine in the AdS/CFT correspondence~\cite{WITTEN,FORCE}.
Spatial Wilson loops map onto two stationary current wires
and reflect on Ampere's law in the gauge theory. Finite
temperature YM theory as well as QCD are known to confine
space-like~\cite{STRINGT} with the string tension proportional
to the magnetic mass $m_M\approx\alpha_s\,T$ asymptotically.
Across $T_c$ the spatial string tension remains about unchanged
from its zero temperature value.

To probe space-like $Q\overline{Q}$ configurations at finite
temperature in AdS/CFT we consider the Euclidean set up
in~\cite{QQ1,QQ2} with a black hole at $z_0=1/\pi\,T$.
The black-hole causes the Euclidean time direction to be compact
with length $\beta=1/T$. The AdS black-hole metric is
\be
ds^2= \frac {R^2}{z^2}
\left(-f dt^2 + {dz^2}/{f}+\sum_i dx_i^2\right)
\label{0}
\ee
with $f=1-z^4/z_0^4$ and $i=1,2,3$.
For large spatial Wilson loops, the relevant part is the near the black
hole and the metric simplifies to be $ds^2=  ({R^2}/{z_0^2})
 (\sum_i dx_i^2)$.
Consider the space-like Wilson lines with one support of length $x$
 running
along the $x^1$-direction  and the other support running at a
distance $y$ and at an angle $\theta$ away from the the $x^1$-direction
in the Euclidean $x^1x^3$-plane. For small $x,y$ the minimal surface
is in general involved, but for $x,y\gg\beta$ it simplifies into
the helicoidal surface~\cite{QQ2}

\be
&&x^1=\tau\,{\rm cos}(\theta\sigma/y)\nonumber\\
&&x^2=\sigma\nonumber\\
&&x^3=\tau\,{\rm sin}(\theta\sigma/y)
\label{1}
\ee
at the bottom attached by two parallel falling surfaces.
Here $\tau, \sigma$ are affine parameters of the 2-dimensional surface,
with $0\leq \tau\leq x$ and $0\leq \sigma\leq y$. For $x\gg y\gg\beta$,
the area swept by the helicoid grazes the black-hole surface.
The straight-down parts are infinite. They
are {\it self-energy} like insertions on the probe lines and will
be subtracted out. In this sense our space-like Wilson lines are
understood as renormalized {\it heavy quarks} or just {\it massless}
light quarks. In both cases, the minimal surface contribution is the
helicoidal surface in the approximatly flat metric near the horizon.

The Nambu-Gotto action action associated to (\ref{1}) is

\be
S(x,y,\theta)=\sigma_S
\int_{0}^{x}\,d\tau\,\int_0^y
d\sigma\,\sqrt{1+\theta^2\frac{{\tau}^2}{y^2}}
\label{2}
\ee
with $\sigma_S$ the spatial string tension,

\be
\sigma_S=\frac 1{2\pi\alpha'} \frac {R^2}{z_0^2}
=\frac{\sqrt{\lambda}}{2\pi}
(\pi\,T)^2=\frac{\pi\sqrt{\lambda}}2\,T^2
\label{3}
\ee
and the AdS radius $R^2=\sqrt{\lambda}\alpha'$.
The AdS/CFT duality implies that

\be
{\rm Tr}\left(W^*(x,y,\theta)e^{-\beta(H-F_0)}\right)
=e^{-\beta F(x,y,\theta)}=e^{-S(x,y,\theta)}
\label{2x}
\ee
with the normalization ${\rm Tr}\,{\bf 1}=1$, and
where $W^*$ is the renormalized spatial Wilson loop and $F$ is
the free energy. The last equality follows from the AdS/CFT duality.
$F(x,y,\theta)$ is the free energy of two stationary
current carrying {\it wires} of length $x$, at a distance $y$, sloped
at an angle $\theta$. Thus

\be
&&F(x,y,\theta)=\nonumber\\
&&{\sigma_S}\,T
\left(xy\sqrt{1+\theta^2\frac{x^2}{y^2}}
+\frac{y^2}{2\theta}{\rm ln}
\frac{+\theta\frac{x}{y}+\sqrt{1+\theta^2\frac{x^2}{y^2}}}
{-\theta\frac{x}{y}+\sqrt{1+\theta^2\frac{x^2}{y^2}}}\right)
\label{5}
\ee
Which gives rise to both a longitudinal and transverse
force.

For $x\gg y$, (\ref{5}) simplifies

\be
F(x\gg y, \theta)={\sigma_S}T\,\theta\,x^2
\label{7}
\ee
which is the area of the surface $x\times \theta x$.
For $\theta=0$, (\ref{5}) reduces to

\be
F(x,y,0)={\sigma_S}T\,xy
\label{6}
\ee
which can be interpreted as the {\it energy} of a fixed
surface $x\times y$ with {\it spatial}
surface tension $\sigma_S\,T$, the natural scale in~\cite{URS}.
(\ref{6}) when reinterpreted in 2+1 dimensions
as an action $\beta\,F$, yields an area law with string
tension $\sigma_S$ as discussed in~\cite{WITTEN,FORCE}.
The forces on the support of the spatial Wilson loop
are
\be
&&F_x/x=-{\sigma_S}T\,y/x\nonumber\\
&&F_y/x=-{\sigma_S}T
\label{AMPERE}
\ee
The former vanishes for $x\gg y$, while the latter
is {\it constant} and {\it attractive} in hot ${\cal N}=4$
SYM.
\footnote{We
are not using the Gibbs relation $E=F-TdF/dT$ for the
spatial Wilson loop as it gives $E=-2F$ nonperturbatively,
and $E=-F$ perturbatively (see below) implying that the
entropy is dominant, a rather unintuitive result.
Our interpretation follows the standard lore for
finite temperature Wilson loops.}

\vskip 0.5cm

{\bf 3. Ampere's Law\,\,.}
The constant force between two infinitly long wires
is a consequence of the fact that Ampere's law confines
in hot ${\cal N}=4$ SYM. The relation of (\ref{AMPERE})
to Ampere's law can be seen in perturbation theory.

The leading cumulant to  (\ref{2x}) is

\be
&&{\rm Tr}\left(W^*(x,y,\theta)e^{-\beta(H-F_0)}\right)\nonumber\\
\approx &&{\rm exp}\left(\frac {\lambda}{2}\,{\rm cos}\theta\,x\,\int\,dr\,
\Delta(\sqrt{r^2+y^2})\right)
\label{2xx}
\ee
with ${\rm Tr}(T^aT^b)=\delta^{ab}/2N_C$ (the Casimir is $N_c/2$) and

\be
\Delta(r)=\frac{T}{4\pi r}\,{\rm coth}(\pi T\,r)
\label{W0}
\ee
the thermal Wightman function for the gluon propagator
in covariant Feynman gauge. In general~\footnote{The ordering
generates a T-independent c-number which is not relevant for
the analysis below.},

\be
&&{\rm Tr}\left(e^{-\beta(H-F_0)}A_\mu^a(x)A_\nu^b(0)\right)=\nonumber\\
&&\delta^{ab}\,\left(g_{44}\Delta_E(x) +g_{ij}\Delta_M(x)\right)
\ee
At large temperature
$\Delta(r)\approx T/4\pi r$, which is checked to
be gauge independent.
(\ref{2xx}) yields the free energy to leading order
in perturbation theory ($x\gg y\gg \beta$)

\be
F(x,y,\theta)=-{\rm cos}\theta\,\frac{\lambda\,T^2}{8\pi}\,x\,{\rm ln}(R/y)
\label{2y}
\ee
where $R$ is a finite cutoff in the relative separation. The
forces stemming from (\ref{2y}) are

\be
\frac{F_x}x=&&-{\rm cos}\theta\,
\frac{\lambda\,T^2}{8\pi}\,\frac{{\rm ln}(y/R)}x\nonumber\\
\frac{F_y}x=&&-{\rm cos}\theta\,\frac{\lambda\,T^2}{8\pi}\,\frac 1y
\label{AMPEREWEAK}
\ee
which are to be compared with (\ref{AMPERE}). For infinite
wires $x\gg y$, only the transverse force survives which
is reminiscent of Ampere's law,

\be
\frac{F_y}{x}=-\frac {N_c/2}{4\pi}\frac {I_1\cdot I_2}{y}
\ee
where $I=g/\beta$ plays the role of a current in the heat bath.
Again, $N_c/2$ is the Casimir.


\vskip 0.5cm

{\bf 4. Non-Static Wilson Loop\,\,.}
These arguments generalize to finite temperature the
arguments presented in~\cite{SPINSPIN}. Indeed,
2 charged particles moving with velocity $\dot{\vec{x}}_{1,2}$
are described by the following {\it non-static Wilson loop}

\be
{\rm Tr}\Big[ &&e^{-\beta(H-F_0)}
{\bf P}\,
{\rm exp}\left(ig\int_{-{\cal T}/2}^{+{\cal T}/2}\,dt_1\,
\,\dot{x}_1\cdot A (x_1)\right)   \nonumber\\
&&\times
{\rm exp}\left(-ig\int_{-{\cal T}/2}^{+{\cal T}/2}
\,dt_2\,\dot{x}_2\cdot A (x_2)
\right)\Big]\,\,.
\label{corx}
\ee
with ${\cal T}$ a large time (in the center of mass).
The first cumulant expansion is driven by the kernel

\be
{\rm Tr}\,
\Bigl[e^{-\beta(H-F_0)}\left(ig\,\dot{x}_1\cdot A (x_1)\right)
\times \left(-ig\,\dot{x}_2\cdot A (x_2)\right)\Bigr]\,\,,
\label{ex}
\ee
which is of the form

\be
\frac{\lambda}2\left(\Delta_E(x_1-x_2)+
\dot{\vec{x}_1}\cdot\dot{\vec{x}_2}\,\Delta_M(x_1-x_2)\right)
\label{EM}
\ee
for the electric and magnetic contribution respectively.

At zero temperature, $\Delta_E=\Delta_M=1/(4\pi^2 x^2)$
after rotation to Euclidean space for regularization,
and (\ref{EM}) becomes

\be
\frac {\lambda}{8\pi^2}\,\frac{1+\dot{\vec{x}_1}\cdot\dot{\vec{x}_2}}
{(t_1-t_2)^2+(\vec{x}_1-\vec{x}_2)^2}
\label{AA}
\ee
The velocity
dependent part translates in Minkowski space to $(1-\vec{v}_1\cdot
\vec{v}_2)$ which is the expected correction to Coulomb's law from
charge motion. This part renormalizes
the Coulomb strength through~\cite{SPINSPIN,BROWN}

\be
\lambda\rightarrow \lambda\,(1-\vec{v}_1\cdot \vec{v}_2)
\label{VV}
\ee
For paralell moving particles with constant
velocity ${\dot{x}}_1={\dot{x}}_2={\dot{x}}$, the potential
$E(L)$ follows from the large time
exponential behaviour of (\ref{corx})

\be
E(L)=&&-\frac{\lambda}{8\pi^2}
\int\,dt \frac {1+{\dot{x}^2}}{(1+{\dot{x}}^2)\,t^2+L^2}
\nonumber\\=&&-\frac{\lambda}{8\pi} \frac 1{L}\,\gamma(1+{\dot{x}^2})
\label{ZERO}
\ee
which is the {\it boosted} form of (Ampere's corrected) Coulomb's law.
Again the extra factor of $1/2$ relates to the Casimir
$N_c/2$. The occurence
of $\gamma=1/\sqrt{1+\dot{x}^2}$ is garenteed by relativity even in the
interacting case, as the dependence in the integral is generically
of the form $(1+\dot{x}^2)\,t^2$ by Lorentz invariance. Thus the
change of variable and the overall factor $\gamma$. So, the induced
magnetic part is necessary even in Coulomb's law when motion is
involved even between very heavy particles.

At finite temperature,
the thermal but free Minkowski contributions
to the Wightman functions (temporal and spatial)
are equal $\Delta_E=\Delta_M=\Delta$
with (modulo regularization)

\be
&&\Delta (t, |x|)=
\nonumber\\&&\frac{T}{8\pi |x|}
\left({\rm coth}(\pi T(|x|+t))+{\rm coth}(\pi T(|x|-t))\right)
\label{W1}
\ee
Here $t=t_1-t_2$ and $|x|=|\vec{x}_1-\vec{x}_2|$,
and (\ref{W1}) reduces to (\ref{W0}) for $t=0$.
For two parallel particles, $|x|=\sqrt{L^2+v^2t^2}$
and at finite temperature the energy (\ref{ZERO}) is now

\be
&&E(L)=-\frac{\lambda T}{8\pi} \nonumber\\
&&\times\,\int\,dt\,\frac {(1-v^2)}{\sqrt{1+v^2t^2}}
{{\rm coth}(\pi\,TL\,(\sqrt{1+v^2t^2}-t))}
\label{EM1}
\ee
Lorentz invariance is more subtle in this case, but can
be checked to follow readily in the zero temperature limit
of (\ref{EM1}).

\vskip 0.5cm

{\bf 5. Ladder Resummation\,\,.}
(\ref{EM1}) receives electric and
magnetic contributions through the Wightman functions

\be
E_E(L)=&&-\frac{\lambda\,}2\int\,dt\,\Delta_E(t,\sqrt{L^2+v^2t^2})\nonumber\\
E_M(L)=&&+\frac{v^2\lambda\,}2\int\,dt\,\Delta_M(t,\sqrt{L^2+v^2t^2})
\label{EM2}
\ee
In the interacting case $\Delta_E\neq \Delta_M$, and also
higher cumulants contribute. At finite temperature, the pure Coulomb 
contribution $\Delta_E$ is screened in strong coupling~\cite{YEE}, 
while the magnetic contribution is not as discussed in (\ref{6}).

To see how this may develope we apply the ordered ladder resummation
to the kernel (\ref{EM},\ref{W1}) for the unscreened magnetic contribution
following the arguments developed in~\cite{ZAREMBO}. Indeed, a rerun of
their arguments yields 

\be
E_{\rm ladd}(L)=\sqrt{v^2}\sigma_S\,L\,F(\pi\,TL)
\label{LADDER}
\ee
with $F(x)=\sqrt{{\rm coth}(x)/2x^3}$.
Only the repulsive contribution to the Bethe-Salpeter kernel
for long times was retained. Noticeably, $v^2\lambda$ transmutes to
$\sqrt{v^2\lambda}$, much like $\lambda$ transmutes to $\sqrt{\lambda}$
in the electric part at strong coupling for ordered ladders~\cite{ZAREMBO}.
At zero temperature, (\ref{LADDER}) reduces to

\be
\sqrt{v^2} \frac{\sqrt{\lambda/2}}{\pi\, L}
\label{ZEROX}
\ee
which is the leading magnetic contribution to the strongly coupled version
of Coulomb's law (\ref{ZERO}).
Note that $F(x\gg 1)\approx 1/x^{3/2}$, which makes (\ref{LADDER}) not
confining but weaker than Coulomb at high temperature.

\vskip 0.5cm

{\bf 6. String Tension and Energy Loss \,\,}
The energy loss reported recently in hot ${\cal N}=4$
SYM in~\cite{YAFFE,DEREK,GUBSER,HERZOG,CAS,VAZ} may
have its origin in the magnetic sector. To support this,
consider an open string of lenght $L$ near the black-hole
horizon $z_0$. When pulled with velocity $v$ along the
x-direction, the string action is

\be
 S=-{L\over2\pi z_0^2}\int dt \sqrt{\lambda\,(1-(z/z_0)^4-v^2)}.
\label{STRINGX}
\ee
A number of remarks follow from this relation:

 \begin{itemize}
   \item The argument of the square root in (\ref{STRINGX}) is
         the string analogue of (\ref{VV}) at finite temperature.
         As $z\rightarrow z_0$ the {\it electric part} 
         $1-(z/z_0)^4$ is completely screened, while the 
         {\it magnetic part} is unchanged. As a result 
         the string action is purely imaginary

         \be 
         S= iv\sigma_S Lt \label{item}. 
         \ee

   \item We {\it interpret the imaginary action} as the 
         energy lost to the black hole horizon as we
         pull the string. The energy loss per length $L$
         is

         \be 
         {E/L}=v\sigma_S, \label{VV3} 
         \ee
         The extra $\gamma$ factor can only be obtained
         through a full analysis with the help of $dz/dx$
         (the tilting of the string in the z-direction 
          for the minimal surface). Our simplified 
          argument is only aimed at understanding the 
          magnetic origin of the loss.

    \item The original analysis in~\cite{YAFFE} shows that
          the energy flows along the dragging string and is
          stored in the string tail infinitesimally above
          $z_0$. Our qualitative description shows that the
          string loses energy to the black hole when it 
          touches its horizon at $z_0$.

    \item Finally, we note that if the above string connects
          a heavy $Q\bar Q$ configuration on the boundary
          (instead of sitting on the black hole horizon), 
          the string action (\ref{STRINGX}) shows that the
          the string will not touch the horizon $z_0$, but 
          instead will break at fixed height

          \be
           z_{max}/z_0= (1-v^2)^{1/4}.
          \ee
         This height translates to a maximum separation
         $l_*=c_1 z_{max},$ for some constant $c_1$. For
         $v=0$, $z_{max}=z_0$ and $c_1$ was determined
         in~\cite{YEE}. The breaking occurs when  
         the screened electric force is
         about to compensate the magnetic force.

 \end{itemize}

%

(\ref{VV3}) for small $v$ is Stoke's law $F=-6\pi\,r\eta\,v$
with a vanishing {\it effective} particle size $r\approx \sigma_S/\eta\approx
\sqrt{\lambda}/(N^2_cT)$. This may suggest that the Ohmic trail
behind a dragged heavy quark is parametrically thin in strong coupling.
Also the Einsten formulae implies that
the diffusion constant of slow particles is $D=T/\sigma_S$
as in~\cite{DEREK}. The relevance of Stoke's law
and the Einstein formulae for moving particles in AdS/CFT were first
suggested in~\cite{GELLMANN} (section Vb) to (re)derive bounds on 
the drag viscosity and the diffusion constant both at finite temperature 
and density.

At strong coupling, a measurement of the
energy loss is a measurement of the
spatial string tension in the underlying gauge theory.
Some of these observations maybe pertinent for QCD and therefore
RHIC. Indeed, QCD confines space-like at all temperatures with 
a spatial string tension $\sigma_S\approx \alpha_s^2\,T^2$ at
asymptotic temperatures.

%

\vskip 0.5cm
{\bf 7. Conclusions\,\,}
We have discussed Ampere's law in hot ${\cal N}=4$ SYM,
for space-like Wilson loops with arbitrary angle $\theta$. The
loops show space-like confinement at all temperatures,
a property shared by other gauge theories. The spatial string
tension is $\sigma_S=\pi\sqrt{\lambda}T^2/2$.
Coulomb's law is screened but Ampere's law is not.

All the results reported
in~\cite{YAFFE,DEREK,GUBSER,HERZOG,CAS,VAZ,URS,URSCRITIC}
involve $\sigma_S$. The technical  reason is due to the
fact that  in the stationary limit (long times, long distances) the subtracted
AdS/CFT surface is mostly determined by the nearly flat metric near
the horizon. In a way, the subtracted surfaces stemming from heavy-quarks
or eikonalized quarks at the {\it top} of the AdS space,
are just massless quarks at the {\it bottom}.  Time is frozen
on the black-hole surface, making most physics spatial.
The physical reason is due to the fact that
at least at high temperature, the gluinos (energy $\pi T$)
together with the scalars (energy $\sqrt{\lambda}T$) decouple
and the electric fields are screened. So very hot ${\cal N}=4$ is
mostly a theory of magnetic gluons. 

In pure lattice YM  the spatial string tension has
been measured~\cite{STRINGT}.
The spatial string tension across $T_c$ shows very little change from
its zero temperature value, in contrast to the temporal string tension.
In zero temperature QCD the temporal
string tension breaks due to $q\overline{q}$
production.
In matter, the {\it static} temporal string tension is Debye-like screened,
however the moving temporal string tension and the
spatial string tension survives at high temperature since
the quarks decouple (energy of order $n\pi T$) space-like. It is of the
order of the magnetic mass squared $m_M^2\approx \alpha_s^2\,T^2$. In
the sQGP at RHIC with $T\approx 1.5\,T_c$, we have
$\sigma_S\approx 1\,{\rm GeV}/{\rm fm}$. Jet quenching at RHIC
may measure this number and its slow dependence on temperature
as suggested by lattice measurements~\cite{STRINGT} across $T_c$.

Finally and rather remarkably, the little bang at RHIC maybe directly mappable
onto a cosmological-like big-bang by the AdS/CFT duality in hot
${\cal N}=4$ SYM~\cite{BIGBANG}, bringing more insights to issues
of transport and dynamics of relevance to the sQGP at RHIC.

\vskip .5cm

{\bf Acknowledgments.\,\,}
IZ thanks Edward Shuryak for a discussion.
The work of IZ was partially supported by the US-DOE grants DE-FG02-88ER40388
and DE-FG03-97ER4014. The work of SJS was supported by KOSEF Grant
R01-2004-000-10520-0 and by SRC Program of the KOSEF with grant
number R11 - 2005- 021.


\end{narrowtext}
\end{document}